\documentclass[12pt]{article}

\begin{document}

\markboth{G. V. Vereshchagin}
{Physical constants  and the Gurzadyan-Xue formula for the dark energy}


\title{Physical constants  and the Gurzadyan-Xue formula for the dark energy}

\author{G. V. Vereshchagin \\
{\it \small International Center for Relativistic Astrophysics,} \\
{\it \small University of Rome "La Sapienza", Physics Department,} \\
{\it \small P.le A. Moro 5, 00185 Rome, Italy.} \\
{\small veresh@icra.it}}


\date{}

\maketitle

\begin{abstract}
We consider cosmological implications of the formula for the dark energy density derived by 
Gurzadyan and Xue\cite{GX,GX2} which predicts a value fitting the observational one. Cosmological models with varying by time physical constants, namely, speed of light and gravitational constant and/or their combinations, are considered. In one of the models, for example, vacuum energy density induces effective negative curvature, while another one has an unusual asymptotic. This analysis also explicitely rises the issue of the meaning and content of physical units and constants in cosmological context.

\end{abstract}


Current observations of supernovae\cite{SNIa,SNIa2,SNIanew} indicate accelerated expansion of the Universe and support models with non-zero cosmological constant $\Lambda$\footnote{Throughout the paper $\Lambda$ is in inverse squared seconds units.}. The idea to interpret the cosmological term as a vacuum energy density $\epsilon_\Lambda=\Lambda c^2/(8\pi G)$ belongs to Zeldovich\cite{zelvac}. However straightforward application of this idea meets certain difficulties. In the four-dimensional flat spacetime $R^4$ and the continuous spectrum of free quantum fields with a ultraviolet cutoff at the Planck scale, the vacuum fluctuation energy density is of the order of $\Lambda_p^4\simeq 10^{76}$GeV$^4$ while the observational data suggest the value $10^{-47}$GeV$^4$. This difference of 123 orders of magnitude constitutes the problem of the cosmological constant\cite{weinvac}.

Gurzadyan and Xue\cite{GX,GX2} have shown that, if one takes into account not all but only \emph{relevant i.e. l=0 modes} for vacuum fluctuations, then one arrives at the formula for
vacuum energy density depending on the fundamental constants and the scale factor of the Universe 

\begin{eqnarray}
\label{rhoLambda}
\rho_{GX}=\frac{\pi}{4}\,\frac{\hbar c}{L_p^2}\,\frac{1}{a^2}=\frac{\pi}{4}\,\frac{c^4}{G}\,\frac{1}{a^2},
\end{eqnarray}
where $\hbar$ is the Planck's constant and the Planck's length is $L_p=\left(\frac{\hbar G}{c^3}\right)^\frac{1}{2}.$ Taking for the scale factor $a$ the Hubble length $l_H=c/H_0$ where $H_0$ is the present value of the Hubble parameter, we find
\begin{eqnarray}
\label{OmegaLambda}
\tilde\Omega_\Lambda=\frac{\rho_\Lambda}{\rho_0 c^2}=\frac{8\pi G}{3H_0^2 c^2}\,\frac{\pi}{4}\,\frac{c^4}{G}\,\frac{H_0^2}{c^2}=\frac{2\pi^2}{3}\simeq 6.58,
\end{eqnarray}
i.e. far less discrepancy than the one mentioned above. It is not difficult to reconcile this discrepancy with observational data ($\Omega^{obs}_\Lambda\simeq 0.7$) simply assuming that the present day scale factor is not given by $l_H$ but is slightly larger and/or allowing various topologies of the Universe.

This formula has the following consequences. If (\ref{rhoLambda}) is considered as a cosmological constant, i.e. $\Lambda$-term
\begin{eqnarray}
\label{Lambda}
\Lambda=\frac{8\pi G \rho_{GX}}{c^2}=\frac{8\pi G}{c^2}\frac{\pi}{4}\,\frac{c^4}{G}\,\frac{1}{a^2}=2\pi^2\left(\frac{c}{a}\right)^2\equiv\textrm{const}.
\end{eqnarray}
then this implies the variation of the speed of light
\begin{eqnarray}
\label{varc}
c(t)=\left(\frac{\Lambda}{2\pi^2}\right)^\frac{1}{2}a(t).
\end{eqnarray}

However, one can also consider the variation of other physical constants in the spirit of Dirac's approach. The aim of this letter is to explore cosmological models with Gurzadyan-Xue formula assuming the presence of some fundamental constant physical quantity and allowing variation of such quantities as the speed of light and the gravitational constant.

Then, from (\ref{rhoLambda}) one has for density parameter
\begin{eqnarray}
\label{OmegaL}
\Omega_\Lambda\equiv\frac{\Lambda}{3H_0^2}=\frac{2\pi^2}{3}\left(\frac{1}{H_0}\right)^2\left(\frac{c}{a}\right)^2,
\end{eqnarray}
with the function $c(a)$. Clearly it does not change with time only if $c\propto a$.

The following cases are possible:

{\bf Case 1.} Neither $c$ nor $G$ vary with time, but $\Lambda\neq$const. The Friedmann equation now has the form
\begin{eqnarray}
H^2+\frac{k' c^2}{a^2}=\frac{8\pi G}{3}\rho, \quad\quad \textrm{where} \quad\quad k'=k-2\pi^2/3.
\end{eqnarray}
Therefore, independently from the value of $k$, the vacuum energy density effectively induces negative curvature. 

{\bf Case 2.}
To keep the vacuum energy density as a cosmological term, as mentioned above, one has to admit varying speed of light.

Then, it follows
\begin{eqnarray}
H^2-\frac{\Lambda'}{3}=\frac{8\pi G}{3}\rho, \quad\quad \textrm{where} \quad\quad \Lambda'=\Lambda\left[1-\frac{3k}{2\pi^2}\right].
\end{eqnarray}
With $k=0,\pm1$ one has modified \emph{positive} cosmological constant at the 
\emph{spatially flat} background.

{\bf Case 3.}
The GX-density of vacuum $\mu_{GX}$ is time invariant
\begin{eqnarray}
\mu_{GX}\equiv\frac{\rho_{GX}}{c^2}=\frac{\Lambda}{8\pi G}=
\frac{\pi}{4}\,\frac{c^2}{G}\,\frac{1}{a^2}=\textrm{const}.
\end{eqnarray}
Then, either one has varying gravitational constant or the speed of light or both. 
We will restrict ourselves with only one of them: $G(a)$ and $c(a)$.

{\bf Case 3.1.}
Substituting the varying gravitational constant
\begin{eqnarray}
	G(a)=\frac{\pi}{4\mu_{GX}}\,\frac{c^2}{a^2}.
\end{eqnarray}
into the Friedmann equation we obtain
\begin{eqnarray}
H^2+\frac{c^2}{a^2}\left[k-\frac{2\pi^2}{3}\left(1+\frac{\rho}{\mu_{GX}}\right)\right]=0,
\end{eqnarray}
where $k=0,\pm1$ is the original curvature.

{\bf Case 3.2.}
For varying speed of light we arrive at
\begin{eqnarray}
	c(a)=2\left(\frac{G\mu_{GX}}{\pi}\right)^\frac{1}{2}a.
\end{eqnarray}
and
\begin{eqnarray}
H^2=\frac{8\pi G}{3}\left[\rho+\mu_{GX}\left(1-\frac{3k}{2\pi^2}\right)\right],
\end{eqnarray}
i.e. again to the case of an effective cosmological constant
\begin{eqnarray}
\Lambda'=8\pi G\mu_{GX}\left[1-\frac{3k}{2\pi^2}\right]=\textrm{const},
\end{eqnarray}
which is equivalent to the case 2 with identification $\Lambda=8\pi G \mu_{GX}$.

{\bf Case 4.}
Assume now that the fundamental constant is the energy density of the vacuum $\rho_{GX}$:
\begin{eqnarray}
\rho_{GX}=\frac{\Lambda c^2}{8\pi G}=\frac{\pi}{4}\,\frac{c^4}{G}\,\frac{1}{a^2}=\textrm{const}.
\end{eqnarray}
Again, there are two cases, $G(a)$ and $c(a)$.

{\bf Case 4.1.}
The gravitational constant now varies
\begin{eqnarray}
G(a)=\frac{\pi}{4\rho_{GX}}\,\frac{c^4}{a^2}.
\end{eqnarray}
Then redefining $\mu_{GX}\equiv\rho_{GX}/c^2$, we come back to the case 3.1.

{\bf Case 4.2.}
The speed of light is varying while the gravitational constant together with the vacuum energy density are fixed. Then
\begin{eqnarray}
c(a)=\left(\frac{4G\rho_{GX}}{\pi}\right)^{1/4}a^{1/2}.
\end{eqnarray}
and
\begin{eqnarray}
H^2=\frac{8\pi G}{3}\rho+\frac{\alpha}{a}.
\end{eqnarray}
where
\begin{eqnarray}
\alpha\equiv\left[\frac{2\pi^2}{3}-k\right]\left(4G\rho_{GX}\right)^{\frac{1}{2}},
\end{eqnarray}

{\bf Case 5.}
Finally, there is always possibility that both constants entering (\ref{rhoLambda}) depend on time, i.e. the speed of light together with the gravitational constant change. Clearly, if they change in combination such that $c^4/(Ga^2)$ remains constant we are back to the case 4. Similarly with constant combination $c^2/(Ga^2)$ we come back to 3. Nothing definite can be said about much larger set of other possibilities, unless the functional dependence of the speed of light and the gravitational constant on time is specified.

Dynamics of Friedmann cosmological models is governed by the system of cosmological equations and the continuity equation, following from Einstein equations. Variation of fundamental constants violates energy conservation and changes the continuity equation\cite{albmag}. This complication does not allow to find analytical solution for most cases considered in this paper. Detailed analysis of the cosmological dynamics and comparison with observational constraints is performed elsewhere\cite{veryeg}.

\vspace{0.2in}

Gurzadyan-Xue formula (1) for the vacuum energy appears to have interesting cosmological implications. Assuming varying speed of light or gravitational constant we analyzed various cosmologies coming out of this formula: with constant $\Lambda$-term (case 2), and of density of the vacuum (case 3) and the vacuum energy density (case 4) and arrived at a bunch of models:
i) spatially flat $\Lambda$-dominated (2, 3.2), ii) curvature-dominated (1), iii) non-Friedmannian (3.1, 4.1), iv) spatially flat $\Lambda=0$ model with dark energy.

It appears that the models depend crucially on which physical constants in particular vary with time, as in the approaches by Dirac (gravitational constant)\cite{dirac,dirac2} Gamow (charge)\cite{gamow}, the speed of light and fine structure constant, recently actively discussed in the cosmological context\cite{webb,barrow,magrev,albmag}. In more general sense, this is directly related with the very content of the physical constants and units, as discussed by Okun\cite{okun}.

\section*{Acknowledgments}
I'd like to thank an anonymous referee for useful remarks, in particular, about the completeness of classification.

\end{document}